\documentclass{article}
\usepackage{amsfonts}
\usepackage{amsxtra}

\usepackage{newsprocl}

\begin{document}

\title{LESS (PRECISION) IS MORE (INFORMATION): QUANTUM INFORMATION IN FUZZY
PROBABILITY THEORY}
\author{P. BUSCH}
\maketitle

\begin{abstract}
A comparison of structural features of quantum and classical physical theories,
such as the information capacity of systems subject to these theories, requires
a common formal framework for the presentation of corresponding concepts (such
as states, observables, probability, entropy). Such a framework is provided by
the notion of statistical model developed in the convexity approach to
statistical physical theories. Here we use statistical models to classify and
survey all possible types of embedding and extension of quantum probabilistic
theories subject to certain reasonable constraints. It will be shown that the
so-called canonical classical extension of quantum mechanics is essentially the
only `good' representation of the quantum statistical model in a classical
framework. All quantum observables are thus identified as fuzzy classical
random variables.
\end{abstract}

\address{Department of Mathematics, University of Hull,\\
  Hull HU6 7RX, England\\
  E-mail: p.busch@hull.ac.uk}

\section{Introduction}

Every physical theory can be formulated as a probabilistic theory: each type
of physical system is characterized through a set of \emph{observables},
representing the possible measurements that can be performed on the system,
with an associated set of \emph{states}, assigning probabilities to the
measurement outcomes. The duality of states and observables is formalized in
the notion of a \emph{statistical duality}, also called \emph{convexity model%
} or \emph{statistical model}, of which quantum mechanics and classical
statistical mechanics are particular realizations. This structure has been
used as the starting point for a reconstruction of quantum mechanics as a
statistical physical theory in fundamental studies in the 1960s-70s\cite
{Dav,Hol,Lud}.

In the past decade, statistical models have been the basis for a series of
very original and penetrating investigations by the late S\l awek Bugajski
and his collaborators (Enrico Beltrametti and Werner Stulpe). They studied a
particular classical representation of quantum mechanical density operators
and observables in terms of probability measures and functions over a `phase
space' which is identified with the set of pure quantum states. This
representation was first formulated by B.~Misra \cite{Misra} and others in
the 1970s, but its significance as `the' \emph{canonical classical extension}
of quantum mechanics was recognized by S.\thinspace Bugajski \cite{Bug1}. E.
Beltrametti and S. Bugajski exhibited some intriguing features of the
canonical classical extension, notably among them the so-called Bell
phenomenon in a classical framework\cite{Bug2} and the fact that \emph{all}
quantum observables (positive operator valued measures, or POVMs), whether
sharp (projection valued measures, PVMs) or not, are represented classically
as \emph{fuzzy random variables}. These observations led to the inception
and development of a generalized classical probability theory, referred to
as fuzzy probability theory\cite{Bug3,Bug4,Gud}.

This line of research was a demonstration of S\l awek Bugajski's creativity
as a researcher in mathematical physics and the foundations of quantum
mechanics. His late work, abruptly cut short by his untimely death in March
2003, comprises a legacy full of intriguing ideas and questions for future
research waiting to be carried further\cite{obit}. S\l awek had hoped to
demonstrate that the formalisms of statistical models and fuzzy probability
theory are useful tools for investigations in quantum information science,
particularly in quantum computation. The aim of the present contribution is
to give a brief exposition of this approach to quantum mechanics and to help
build an intuitive understanding of the intriguing features referred to
above. I will review the concept of statistical models, recall the quantum
and classical examples, and provide a comprehensive account of all possible
mutual embedding schemes. I will show that the canonical classical extension
of quantum mechanics is the only `good' classical representation of quantum
mechanics as a statistical model. Since quantum observables are in this
description represented as fuzzy classical random variables, it follows that
the enhanced information storage and cryptographic capacities of quantum
systems must be seen as a consequence of the fact that the set of quantum
observables appears as a restriction of the set of all classical random
variables to a subset of fuzzy random variables.

The title phrase, ``Less is more...'', refers to the fact that quantum
uncertainty can be utilized as resources for information processing purposes in
the broadest sense. For example, noncommuting pairs of observables can be
measured jointly if one allows for an appropriate pay-off between the degrees
of unsharpness in the respective marginal measurements, quantified by suitable
uncertainty relations. Another example is given by the existence
of \emph{informationally complete} POVMs, which are necessarily \emph{unsharp%
} observables. Further, it has been observed that certain state
discrimination procedures are optimally performed by POVMs which are not
PVMs. Finally, measurements which are \emph{less} precise may be designed so
as to be less invasive, thus allowing \emph{more }control over the observed
system. Incidentally, Th.\thinspace Konrad chose the same motto for his
doctoral thesis on unsharp quantum measurements\cite{Kon}. While our choices
were made independently of each other, we may both have been (consciously or
subconsciously) influenced by H. Dehmelt's famous review \cite{Dehm} which
was written in the same spirit.

The reader may wish to read the present article in conjunction with the related
contributions by Chris Fuchs\cite{Fuchs} and Guido Bacchiagaluppi in this
volume. The former uses the affine classical embedding of quantum states
induced by an informationally complete POVM to explore aspects of the concept
of quantum information. The latter investigates the description of state
changes in the framework of the Misra-Bugajski canonical classical extension.

This work is dedicated to the memory of S\l awek Bugajski.

\section{Statistical models}

Every type of physical system can be prepared and observed in a variety of
ways. In order to exhibit regularities in the observation data, it must be
possible to reproduce (practically) the `same' conditions of preparation and to
carry out operations which can be identified as (practically) the `same'
measurements, so as to be able to determine that different preparations tend to
lead to different measurement outcomes. Thus, the theoretical modelling of a
physical system should be based on the structures of a set $\mathcal{S}$ of
\emph{states}, which represent the different possible, reproducible
preparations, and a set $\mathcal{O}$ of \emph{observables}, which represent
the different possible measurements. A state and an observable should together
determine a probability distribution that corresponds to the expected
frequencies of the outcomes of the same (class of) measurement, repeated many
times on a system (or a collection of independent systems) prepared in the same
state.

These ideas entail some basic properties of the sets $\mathcal{S}$ and $%
\mathcal{O}$ which we will spell out next. The possibility of randomly choosing
different states, with fixed probability weights, and subjecting them to
\emph{independent} runs of the same measurement implies that the set of states
should be convex, thus including all mixtures of any finite or (by
extrapolation) countable subsets. Moreover it follows that each measurement
outcome, labelled by a value $a_{k}$ ($k=1,2,\dots $), say, gives rise to an
affine map
\begin{equation}
E_{k}:\mathcal{S}\rightarrow \lbrack 0,1]\,,\;\;\rho \mapsto E_{k}\left(
\rho \right) \equiv p_{\rho }\left( a_{k}\right)  \label{effect1}
\end{equation}
where $p_{\rho }\left( a_{k}\right) $ is the probability of the outcome $%
a_{k}$ in the state $\rho \in \mathcal{S}$. We can ensure that every
measurement is certain to produce an outcome by allowing for a null event, $%
a_{0}$, to represent the situation where there is no response. It follows
that the sum of probabilities $\sum_{k}p_{\rho }\left( a_{k}\right) =1$ for
all states $\rho $. We summarize this by writing $\sum_{k}E_{k}=I$, where $%
I:\rho \mapsto 1$ is the constant unit map that sends every state to the
number 1. The totality of all maps $E_{k}$ defines the observable measured
in the experiment under consideration.

This consideration generalizes as follows. The set of states of a given type
of physical system is modelled as a convex set $\mathcal{S}$. The set $%
\mathcal{O}$ of observables is given by a set of affine maps
\begin{equation}
A:\mathcal{S}\rightarrow M\left( \Omega \right) _{1}^{+}\,,\;\;\rho \mapsto
A\left( \rho \right) \equiv \mu _{\rho }^{A}  \label{obs1}
\end{equation}
from $\mathcal{S}$ to the convex set $M\left( \Omega \right) _{1}^{+}$ of
probability measures over some measurable spaces $\left( \Omega ,\Sigma
\right) $, where $\Omega $ denotes the set of measurement outcomes and $%
\Sigma $ a $\sigma $-algebra of subsets of $\Omega $ appropriate to each
measurement represented in the model.

Each observable $A$, together with a subset $X\in \Sigma $, gives rise to an
affine map
\begin{equation}
E\left( X\right) :\mathcal{S}\rightarrow \lbrack 0,1]\,,\;\;\rho \mapsto
E\left( X\right) \left( \rho \right) =:A\left( \rho \right) \left( X\right)
\equiv E_{\rho }\left( X\right) .  \label{effect2}
\end{equation}
Any affine map $a:\mathcal{S}\rightarrow \lbrack 0,1]$ is called an \emph{%
effect}. The map
\begin{equation}
\Sigma \ni X\mapsto E\left( X\right)   \label{effmeas}
\end{equation}
has the properties of a measure, with normalization $E\left( \Omega \right)
\equiv I$, the property $E\left( \emptyset \right) =O$ (=the constant zero
map which is itself an effect), and $\sigma $-additivity following from that
of all the probability measures $E_{\rho }$. We see that any observable $A$
gives rise to an \emph{effect valued measure }$E\equiv E^{A}$. Conversely,
every effect valued measure $E$ on some measurable space $\left( \Omega
,\Sigma \right) $ induces an observable $A\equiv A^{E}$ as an affine map
from $\mathcal{S}$ to $M\left( \Omega \right) _{1}^{+}$. We have thus, in
effect, two equivalent definitions of an observable.

Given the set $\mathcal{O}$ of observables, we can define the set of all
physically realizable effects, denoted $\mathcal{E}$, as the union of the
ranges of all the effect valued measures associated with the observables in $%
\mathcal{O}$. This set $\mathcal{E}$ is thus a subset of the set $\mathcal{E}%
\left( \mathcal{S}\right) $ of \emph{all} effects on $\mathcal{S}$. The set $%
\mathcal{E}$ inherits a natural partial order defined in $\mathcal{E}\left(
\mathcal{S}\right) $ as follows:
\begin{equation}
a\leq b\;\mathrm{if\;and\;only\;if\;}a\left( \rho \right) \leq b\left( \rho
\right) \;\mathrm{for\;all}\;\rho \in \mathcal{S}.  \label{order}
\end{equation}
We can thus write $\mathcal{E}\left( \mathcal{S}\right) =\left[ O,I\right] $%
. The set $\mathcal{E}$ represents the collection of all \emph{simple}
observables, that is, those with only two outcomes (also called yes-no
observables): for $a\in \mathcal{E}$, we also have $a^{\prime }:=I-a\in
\mathcal{E}$, and the map $\left\{ 1\right\} \mapsto a$, $\left\{ 0\right\}
\mapsto a^{\prime }$ defines an effect valued measure on (the power set of) $%
\Omega =\left\{ 0,1\right\} $. We can take $\mathcal{O}$ to be closed under
all coarse-grainings of its observables (understood as restrictions of the
associated effect valued measures on $\left( \Omega ,\Sigma \right) $ to sub-%
$\sigma $-algebras of $\Sigma $.

The map
\begin{equation}
\mathcal{E}\left( \mathcal{S}\right) \ni a\mapsto a^{\prime }=I-a\in
\mathcal{E}\left( \mathcal{S}\right)   \label{complem}
\end{equation}
defines a kind of complement, or negation, and one can consider a definition
of weak orthogonality of effects: $a\perp b$ if $a+b\leq I$. This relation
is not an orthocomplement since one can have pairs of effects satisfying $%
a\perp b$ while both have a common nonzero lower bound, $O\leq c\leq a,b$.
With the partial operation
\begin{equation}
\mathcal{E}\ni a,b\mapsto a\oplus b:=a+b\;\mathrm{if}\;a+b\in \mathcal{E},
\label{partsum}
\end{equation}
the set of effects $\mathcal{E\ }$assumes the structure of an \emph{effect
algebra}. The connection between effect algebras and statistical models has
been thoroughly analyzed by Beltrametti and Bugajski\cite{effalg} and
independently by Gudder.\cite{Gud-effalg}

We will make the assumption that the set of physical effects $\mathcal{E}$
separates $\mathcal{S}$, that is: for any two different $\rho ,\rho ^{\prime
}\in \mathcal{S}$, there is an effect $a\in \mathcal{E}$ such that $a\left(
\rho \right) \neq a\left( \rho ^{\prime }\right) $. This is a natural
assumption and can be satisfied by identifying all states that would
otherwise be indistinguishable. We then summarize these consideration by
defining a \emph{statistical model }as a pair $\langle \mathcal{S},\mathcal{E%
}\rangle $ consisting of a convex set of states and a separating set of
physical effects.

It is convenient to consider convex structures as embedded into their
natural linear extensions. As a convex set, $\mathcal{S}$ is part of a
vector space $\mathcal{V=V}\left( \mathcal{S}\right) $ over the field of
real numbers which we shall consider to be the span of $\mathcal{S}$. The
set of effects $\mathcal{E}\left( \mathcal{S}\right) $ is a convex subset of
the vector space $\mathcal{A}\equiv \mathcal{A}_{b}\left( \mathcal{S}\right)
$ of all (real-valued) bounded affine functionals of $\mathcal{S}$. Let $%
\mathcal{A}^{\ast }$ denote the algebraic dual space to $\mathcal{A}$, that is,
the set of linear functionals on $\mathcal{A}$. Then, for $\rho \in
\mathcal{V}$, the map $a\mapsto a\left( \rho \right) $ is a linear functional
on $\mathcal{A}$, which implies that $\mathcal{V}$ is injectively embedded into
$\mathcal{A}^{\ast }$. Thus we can introduce a (nondegenerate) bilinear form on
$\mathcal{A}^{\ast }\times \mathcal{A}$, $\langle \cdot ,\cdot \rangle $, such
that for $\rho \in \mathcal V$,
\begin{equation}
\langle \rho ,a\rangle =\rho \left( a\right) =a\left( \rho \right) \,.
\label{bilin1}
\end{equation}
With the vector space $\mathcal{W=W}\left( \mathcal{E}\right) $ generated as
the span of $\mathcal{E}$, the pair $\langle \mathcal{V},\mathcal{W}\rangle $%
, equipped with the nondegenerate form $\langle \cdot |\cdot \rangle $, is
referred to as a\emph{\ dual pair}. Every statistical model can thus be
embedded in a dual pair of vector spaces.

The condition that $\mathcal{E}$ separates $\mathcal{S}$ ensures that the
vector space $\mathcal{V}$ can be equipped with a norm whose unit ball is $%
\mathcal{B}:=\mathrm{conv}\left( \mathcal{S\cup \,-S}\right) $; this norm,
called the \emph{base norm}, is the Minkowski functional of $\mathcal{B}$,
defined via
\begin{equation}
m\left( \rho \right) :=\mathrm{\inf }\left\{ \lambda :\rho \in \lambda
\mathcal{B}\right\} =:\left| \left| \rho \right| \right| _{1}.  \label{norm1}
\end{equation}
In this way, $\mathcal{V}$ is a base norm space. Since bounded affine
functionals on $\mathcal{S}$ extend uniquely to bounded linear functionals
on $\mathcal{V}$, $\mathcal{A}_{b}\left( \mathcal{S}\right) $ is a subspace
of the dual space $\mathcal{V}^{\ast }$ of $\mathcal{V}$. The set $\left[ O,I%
\right] =\mathcal{E}\left( \mathcal{S}\right) \subset \mathcal{A}_{b}\left(
\mathcal{S}\right) $ is called \emph{order unit interval}, and it makes $%
\mathcal{A}_{b}\left( \mathcal{S}\right) $ an order unit space with the
order unit norm equal to the dual space norm inherited from $\mathcal{V}%
^{\ast }$ and the norm unit ball given by $\left[ -I,I\right] $. If the
normed vector spaces $\mathcal{V,W}$ are complete, the pair $\langle
\mathcal{V},\mathcal{W}\rangle $ is called a \emph{statistical duality}. A
lucid introduction to these structures and their physical context can be
found in the work of W. Stulpe{}\cite{Stulpe}; for further details, cf. also
Ref.\negthinspace \cite{effalg} and references therein.

\section{Classical and quantum statistical models}

The traditional classical statistical model is determined by a measurable
space $\left( \Gamma ,\mathcal{B}\left( \Gamma \right) \right) $, where $%
\Gamma $ is the state space of a dynamical system and $\mathcal{B}\left(
\Gamma \right) $ is a $\sigma $-algebra of subsets of $\Gamma $. Usually, $%
\Gamma $ is a topological or metric space, e.g. a (subspace of) $\Bbb{R}^{n}$%
, and $\mathcal{B}\left( \Gamma \right) $ will then be the associated Borel
algebra. The set of (statistical) states $\mathcal{S}_{c}$ is usually taken
to be the set of all measures $M_{1}^{+}\left( \Gamma ,\mathcal{B}\left(
\Gamma \right) \right) $ or sometimes the set of (measures with) probability
densities with respect to some reference measure, $L^{1}\left( \Gamma
,dm\right) _{1}^{+}$, such as $L^{1}\left( \Bbb{R}^{n},d^{n}\mathbf{x}%
\right) _{1}^{+}$. Following Beltrametti and Bugajski\cite{effalg}, we define
the set of \emph{measurable effects} $\mathcal{E}_{c}$ as the collection of
affine functionals on $\mathcal{S}_{c}$, determined by some measurable function
$f:\Gamma \rightarrow \left[ 0,1\right] $:
\begin{equation}
a_{f}:\mathcal{S}_{c}\rightarrow \left[ 0,1\right] ,\;\;\mu \mapsto
a_{f}\left( \mu \right) =\int_{\Gamma }f\,d\mu \text{.}  \label{classeff}
\end{equation}
The order relation $a_{f}\leq a_{g}$ is now equivalent to $f\leq g$ in the
case $\mathcal{S}_{c}=M_{1}^{+}\left( \Gamma ,\mathcal{B}\left( \Gamma
\right) \right) $ [take $\mu =\delta _{\gamma }$, the Dirac measure
supported at $\gamma \in \Gamma $] or $f\leq g$ almost everywhere with
respect to the measure $m$ for $\mathcal{S}_{c}=L^{1}\left( \Gamma
,dm\right) _{1}^{+}$.

An observable is an affine map $A:\mathcal{S}_{c}\rightarrow M\left( \Omega
,\Sigma \right) _{1}^{+}$ from the given set of classical states to the set
of probability measures on a measurable space $\left( \Omega ,\Sigma \right)
$ of measurement outcomes. This comprises the traditional definition of a
classical observable as a function on phase space (random variable). Let $%
F:\Gamma \rightarrow \Omega $ be a measurable function, then we define an
affine map $A_{F}:\mathcal{S}_{c}\rightarrow M\left( \Omega ,\Sigma \right)
_{1}^{+}$ as follows:
\begin{equation}
A_{F}\mu \left( X\right) :=\mu \left( F^{-1}\left( X\right) \right)
=\int_{\Gamma }\delta _{F\left( \gamma \right) }\left( X\right) \,d\mu
\left( \gamma \right) \,,\;\;X\in \Sigma \,.  \label{classobs}
\end{equation}
The associated effect valued measure is given by
\begin{equation}
E_{F}\equiv F^{-1}:\Sigma \rightarrow \mathcal{B}\left( \Gamma \right)
\,,\;\;X\mapsto F^{-1}\left( X\right) =a_{\chi _{F^{-1}\left( X\right) }}\,.
\label{classpvm}
\end{equation}
As a function on $\Gamma $, this corresponds to the indicator function $\chi
_{F^{-1}\left( X\right) }$. (Note that $\chi _{F^{-1}\left( X\right) }\left(
\gamma \right) =\delta _{F\left( \gamma \right) }\left( X\right) $, hence
the above integrand is a measurable and indeed integrable function.)

This construction of the standard classical observable extends to the class
of observables with measurable effects as follows. The function $\left(
\gamma ,X\right) \mapsto \chi _{F^{-1}\left( X\right) }\left( \gamma \right)
$ is a particular instance of a Markov kernel
\begin{equation}
K:\Gamma \times \Sigma \rightarrow \left[ 0,1\right] \,,\;\;\left( \gamma
,X\right) \mapsto K\left( \gamma ,X\right) ,  \label{Markovkern}
\end{equation}
that is, a map with the property that $K\left( \gamma ,\cdot \right) \in
M\left( \Omega ,\Sigma \right) _{1}^{+}$ and $K\left( \cdot ,X\right) $ is a
measurable function, in fact, an effect. Then the following defines an
observable $A_{K}$ which represents a \emph{fuzzy random variable}:
\begin{equation}
A_{K}\mu \left( X\right) :=\int_{\Gamma }K\left( \gamma ,X\right) \,d\mu
\left( \gamma \right) \,,\;\;X\in \Sigma \,,  \label{fuzzyrv}
\end{equation}
with associated effect valued measure
\begin{equation}
E_{K}:\Sigma \rightarrow \mathcal{E}_{c}\,,\;\;X\mapsto a_{K\left( \cdot
,X\right) }\,.  \label{classpov}
\end{equation}

A quantum statistical model is usually based on the set of states $\mathcal{S%
}_{q}=\mathcal{S}\left( \mathcal{H}\right) $ given by the density operators $%
\rho $ of a separable complex Hilbert space $\mathcal{H}$. Conventional
quantum mechanics takes its set of physical effects $\mathcal{E}_{q}^{p}$ as
the set of all orthogonal projection operators $P$, $P=P^{\ast }=P^{2}$.
Operational quantum physics is based on the set of all effects $\mathcal{E}%
_{q}=\mathcal{E}\left( \mathcal{H}\right) =\left[ O,I\right] $, given by the
operators $a$ for which $O\leq a\leq I$. In both cases, projections and
effects define affine functionals on $\mathcal{S}_{q}$ via the trace
formula, $a\left( \rho \right) =\mathrm{tr}\left[ \rho \cdot a\right] $.
Both sets $\mathcal{E}_{q}^{p}$ and $\mathcal{E}_{q}$ are effect algebras,
with the set of projections being the set of extreme elements of the convex
set of effects. $\mathcal{E}_{q}^{p}$ is an orthocomplemented non-Boolean
lattice under the ordering $a\leq b$ and complementation map $a\mapsto
a^{\prime }$ inherited from $\mathcal{E}_{q}$.

A characteristic difference between quantum and classical statistical models
is given by the fact that $\mathcal{S}_{c}\mathcal{\ }$is a simplex (all
states have a unique representation as a convex combination (finite,
countable or continuous) in terms of the extreme elements of $\mathcal{S}%
_{c} $, while all mixed quantum states allow many decompositions into pure
states. We illustrate this state of affairs by means of the smallest nontrivial
example of a quantum statistical model associated with the Hilbert space
$\mathcal{H}=\Bbb{C}^{2}$. This will be compared with the classical statistical
model based on a set $\Omega =\left\{ 1,2,3,4\right\}$.

The classical statistical model for $\Omega _{4}=\left\{ 1,2,3,4\right\} $,
with $\Sigma =\mathcal{P}\left( \Omega _{4}\right) $ (the power set of $%
\Omega _{4}$), is given by $\mathcal{S}\left( \Omega _{4}\right) \subset
\Bbb{R}^{4}\equiv \mathcal{V}\left( \Omega _{4}\right) $, where
\begin{equation}
\mathcal{S}\left( \Omega _{4}\right) =\left\{ p=\left(
p_{1},p_{2},p_{3},p_{4}\right) :p_{k}\geq
0,\,p_{1}+p_{2}+p_{3}+p_{4}=1\right\} \,.  \label{class-states-4dim}
\end{equation}
This is a tetrahedron with vertices $\left( 1,0,0,0\right) ,\left(
0,1,0,0\right) ,\left( 0,0,1,0\right) ,\left( 0,0,0,1\right) $. The set of
effects $\mathcal{E}\left( \Omega _{4}\right) $ comprises all the maps
\begin{equation}
a:p\mapsto a\left( p\right) =a\left( \sum p_{k}e_{k}\right) =\sum
p_{k}a\left( e_{k}\right) =\sum p_{k}a_{k}  \label{class-eff-4dim}
\end{equation}
for which $a\left( p\right) \in \left[ 0,1\right] $. [Here the $e_{k}$
denote the Cartesian unit basis vectors, and $a_{k}:=a\left( e_{k}\right) $%
.] This condition entails that $0\leq a_{k}\leq 1$. Hence the set of effects
can be identified as a subset of $\Bbb{R}^{4}\equiv \mathcal{W}\left( \Omega
_{4}\right) $:
\begin{equation}
\mathcal{E}\left( \Omega _{4}\right) =\left\{ a=\left(
a_{1},a_{2},a_{3},a_{4}\right) :0\leq a_{k}\leq 1\right\} \,.
\label{class-eff-4dim2}
\end{equation}
This is a hypercube with the vertices given by the 16 points whose coordinates
are quadruples of 0s and 1s. The bilinear form associated with
the dual pair $\langle \mathcal{V}\left( \Omega _{4}\right) ,\mathcal{W}%
\left( \Omega _{4}\right) \rangle $ is the Euclidean inner product,
\begin{equation}
\mathcal{V}\left( \Omega _{4}\right) \times \mathcal{W}\left( \Omega
_{4}\right) \ni \left( p,a\right) \mapsto \langle p,a\rangle =\sum
p_{k}a_{k}=p\cdot a\,.  \label{class-bilin}
\end{equation}

The quantum statistical model for $\mathcal{H}=\Bbb{C}^{2}$ (a spin-$\frac{1%
}{2}$ system or generally a qubit) can be parameterized in the Cayley
representation as follows. As a basis of the space of 2$\times $2 matrices
(linear operators of $\Bbb{C}^{2}$) we take
\begin{equation}
I=\left(
\begin{array}{cc}
1 & 0 \\
0 & 1
\end{array}
\right) ,\;\;\sigma _{1}=\left(
\begin{array}{cc}
0 & 1 \\
1 & 0
\end{array}
\right) ,\;\;\sigma _{2}=\left(
\begin{array}{cc}
0 & -\iota  \\
\iota  & 0
\end{array}
\right) ,\;\;\sigma _{3}=\left(
\begin{array}{cc}
1 & 0 \\
0 & -1
\end{array}
\right)   \label{Pauli}
\end{equation}
where the $\sigma _{k}$ are the Pauli matrices. Every density operator can
be written as $\rho =\frac{1}{2}\left( r_{0}I+\mathbf{r}\cdot \mathbf{\sigma
}\right) $ [where $\mathbf{r}=\left( r_{1},r_{2},r_{3}\right) \in \Bbb{R}^{3}
$, $\mathbf{\sigma }=\left( \sigma _{1},\sigma _{2},\sigma _{3}\right) $]
and is thus associated with a vector $\tilde{\rho}=\left( r_{0},\mathbf{r}%
\right) \in \Bbb{R}^{4}$. The fact that $\rho $ assigns probability one to the
effect $I$ entails that tr$\left[ \rho \right] =r_{0}=1$. The condition that
$\rho \geq O$ can be expressed in terms of the eigenvalues of $\rho $:
$\frac{1}{2}\left( 1\pm \left| \mathbf{r}\right|
\right) \geq 0$, which is equivalent to the Euclidean norm $\left| \mathbf{r}%
\right| \leq 1$. The set $\mathcal{S}\left( \Bbb{C}^{2}\right) $ is thus
isomorphically represented by a subset $\widetilde{\mathcal{S}}\left( \Bbb{C}%
^{2}\right) $ of $\Bbb{R}^{4}$,
\begin{equation}
\mathcal{S}\left( \Bbb{C}^{2}\right) =\left\{ \rho :\mathrm{tr}\left[ \rho
\right] =1,\,\rho \geq O\right\} \leftrightarrow \widetilde{\mathcal{S}}%
\left( \Bbb{C}^{2}\right) =\left\{ \tilde{\rho}=\left( 1,\mathbf{r}\right)
:\left| \mathbf{r}\right| \leq 1\right\} \,.  \label{q-states-2dim}
\end{equation}
$\widetilde{\mathcal{S}}\left( \Bbb{C}^{2}\right) $ is a 3-dimensional unit
ball embedded in the hyperplane $r_{0}=1$. The effects of $\Bbb{C}^{2}$ are
the operators (matrices) $a=\frac{1}{2}\left( a_{0}I+\mathbf{a}\cdot \mathbf{%
\sigma }\right) $ satisfying the condition $O\leq a\leq I$, which is
equivalent to $0\leq \frac{1}{2}\left( a_{0}\pm \left| \mathbf{a}\right|
\right) \leq 1$. Thus $\mathcal{E}\left( \Bbb{C}^{2}\right) $ is isomorphic
to a subset $\widetilde{\mathcal{E}}\left( \Bbb{C}^{2}\right) $ of $\Bbb{R}%
^{4}$,
\begin{equation}
\mathcal{E}\left( \Bbb{C}^{2}\right) =\left\{ a:O\leq a\leq I\right\}
\leftrightarrow \widetilde{\mathcal{E}}\left( \Bbb{C}^{2}\right) =\left\{
\tilde{a}=\left( a_{0},\mathbf{a}\right) :\tfrac{1}{2}\left( a_{0}\pm \left|
\mathbf{a}\right| \right) \leq 1\right\} \,.  \label{q-eff-2dim}
\end{equation}
$\mathcal{E}\left( \Bbb{C}^{2}\right) $ (or rather $\widetilde{\mathcal{E}}%
\left( \Bbb{C}^{2}\right) $) can be visualized as a convex diamond-shaped
figure that is the convex hull of the `top' and `bottom' vertices given by $%
I\leftrightarrow \left( 2,0\right) $, $O\leftrightarrow \left( 0,\mathbf{0}%
\right) $ and the surface of the sphere $\left\{ \left( 1,\mathbf{a}\right)
:\left| \mathbf{a}\right| =1\right\} $.

The bilinear form associated with the dual pair of vector spaces which host $%
\mathcal{S}\left( \Bbb{C}^{2}\right) $, $\mathcal{E}\left( \Bbb{C}%
^{2}\right) $ is given as follows:
\begin{equation}
\langle \rho ,a\rangle =tr\left[ \rho \cdot a\right] =\tfrac{1}{2}\left(
a_{0}+\mathbf{a}\cdot \mathbf{r}\right) =\tfrac{1}{2}\tilde{a}\cdot \tilde{%
\rho}  \label{q-bilin}
\end{equation}

\section{Embeddings and extensions of statistical models}

We consider next the various possible ways of embedding or extending one
statistical model $\langle \mathcal{S},\mathcal{E}\rangle $ (in)to another, $%
\langle \mathcal{S}^{\prime },\mathcal{E}^{\prime }\rangle $, in a `natural'
way. That is, we are asking for an association of states $\rho $ and effects
$a$ of $\langle \mathcal{S},\mathcal{E}\rangle $ with states $\rho ^{\prime }
$ and effects $a^{\prime }$ of $\langle \mathcal{S}^{\prime },\mathcal{E}%
^{\prime }\rangle $, $\rho ,a\leftrightarrow \rho ^{\prime },a^{\prime }$,
which is such that probabilities are preserved: $\langle \rho ,a\rangle
=\langle \rho ^{\prime },a^{\prime }\rangle $; further we require that the
convex structures are respected by the sought correspondence.

We shall investigate two obvious choices which ensure that all states of $%
\mathcal{S}$ will be represented:

\begin{description}
\item[(A)]  There is an injective affine map $\Phi :\mathcal{S}\rightarrow
\mathcal{S}^{\prime }$; this \emph{embedding map }should give a faithful
representation of the states of $\mathcal{S}$ in a possibly `larger' state
space..

\item[(B)]  There is a surjective affine map $R:\mathcal{S}^{\prime
}\rightarrow \mathcal{S}$; this \emph{reduction map }would describe the
system represented by $\mathcal{S}$, possibly as a part of subsystem of a
larger, compound system.
\end{description}

In fact, both maps could exist simultaneously, being inverses to each other.
In this case the two state spaces are isomorphic. In both cases (A) and (B)
we will have to determine whether all effects of $\mathcal{E}$ can indeed be
represented by effects of $\mathcal{E}^{\prime }$, as envisaged.

Assume an affine map $\Phi :\mathcal{S}\rightarrow \mathcal{S}^{\prime }$ is
given. This extends uniquely to a linear map from $\mathcal{V}\left(
\mathcal{S}\right) $ to $\mathcal{V}\left( \mathcal{S}^{\prime }\right) $,
which we will also denote $\Phi $. This induces the dual map $\Phi ^{\ast }:%
\mathcal{V}\left( \mathcal{S}^{\prime }\right) ^{\ast }\rightarrow \mathcal{V%
}\left( \mathcal{S}\right) ^{\ast }$, defined by the condition
\begin{equation}
\langle \Phi \rho ,a^{\prime }\rangle =\langle \rho ,\Phi ^{\ast }a^{\prime
}\rangle \,,\;\;\mathrm{i.e.},\;\;\Phi ^{\ast }a^{\prime }=a^{\prime }\circ
\Phi \,,\;\;\rho \in \mathcal{V}\left( \mathcal{S}\right) \,,\;a^{\prime
}\in \mathcal{V}\left( \mathcal{S}^{\prime }\right) ^{\ast }\,.
\label{dualmap1}
\end{equation}
$\Phi ^{\ast }$ sends $\mathcal{E}\left( \mathcal{S}^{\prime }\right) $ into
$\mathcal{E}\left( \mathcal{S}\right) $: indeed, for $a^{\prime }\in
\mathcal{E}\left( \mathcal{S}^{\prime }\right) $, we have $\left( \Phi
^{\ast }a^{\prime }\right) \left( \rho \right) =a^{\prime }\left( \Phi \rho
\right) \in \left[ 0,1\right] $ for all $\rho \in \mathcal{S}$, that is, $%
\Phi ^{\ast }a^{\prime }\in \mathcal{E}\left( \mathcal{S}\right) $. The
requirement of injectivity of $\Phi $ implies that the range of $\Phi ^{\ast
}$ is (weak-$\ast $) dense in $\mathcal{V}\left( \mathcal{S}\right) ^{\ast }$%
. However, this does not imply that $\Phi ^{\ast }\left( \mathcal{E}\left(
\mathcal{S}^{\prime }\right) \right) $ is dense in $\mathcal{E}\left(
\mathcal{S}\right) $. We will say that an injective affine map $\Phi :%
\mathcal{S}\rightarrow \mathcal{S}^{\prime }$ induces a \emph{`good'
embedding} of the model $\langle \mathcal{S},\mathcal{E}\rangle $ within $%
\langle \mathcal{S}^{\prime },\mathcal{E}^{\prime }\rangle $ if $\Phi ^{\ast
}\left( \mathcal{E}^{\prime }\right) $ is at least dense in $\mathcal{E}$,
so that every physical effect of the original model can be described by a
physical effect of the extended model. We shall see that there is no `good'
classical embedding of the full quantum statistical model.

Now consider the case where there exists an affine, surjective reduction map
$R:\mathcal{S}^{\prime }\rightarrow \mathcal{S}$. This having a surjective
extension to $\mathcal{V}\left( \mathcal{S}^{\prime }\right) $, it follows
that the dual map $R^{\ast }$ provides an injective representation of all
effects in $\mathcal{E}\left( \mathcal{S}\right) $ as effects in $\mathcal{E}%
\left( \mathcal{S}^{\prime }\right) $ via $R^{\ast }a=a\circ R$. The map $R$
is thus seen to be a \emph{`good' extension} of the statistical duality $%
\langle \mathcal{S},\mathcal{E}\rangle $ to $\langle \mathcal{S}^{\prime },%
\mathcal{E}^{\prime }\rangle $ (provided it exists). We will review the
Misra-Bugajski map as an example of a `good' classical extension of the
quantum statistical model.

\section{Embeddings and extensions of quantum statistical models}

We give an overview of the known types of embeddings and extensions of
quantum statistical models, including the classical representations. We
begin with an `unsuccessful' attempt, one which nevertheless has provided
important insights into the structure of quantum mechanics.

\subsection{Wigner function - a counter example}

The question of classical representations of quantum mechanics is as old as
quantum mechanics itself: already in Born's famous paper of 1926 introducing
the probability interpretation\cite{Born}, the question of extensions of
quantum mechanics in the framework of a classical theory with hidden
variables was raised. A few years later, Wigner\cite{Wig} attempted to
establish a phase space formulation of quantum mechanics and found an
injective (in fact isometric) affine map $W:\mathcal{S}_{q}\rightarrow
L^{1}\left( \Gamma ,dq\,dp\right) $ of the density operators $\rho $ of a
quantum particle to integrable and normalized functions of phase space. This
map yields the correct marginal position and momentum distributions
associated with each $\rho $, but the functions $W\rho \left( q,p\right) $
are \emph{not} nonnegative (apart from a `few' exceptions such as Gaussian
wave functions). That is to say that the Wigner map $W$ does \emph{not} map
into $L^{1}\left( \Gamma ,dq\,dp\right) _{1}^{+}$, the probability densities
on phase space. This rules out $W$ as a classical embedding of quantum
mechanics in the sense defined here.\footnote{%
See also the penetrating remarks of S. Bugajski\cite{Bug1}.}

\subsection{Classical embedding induced by an observable}

However, we have already seen examples of candidates for classical
embeddings, namely, in the form of the affine maps $A:\mathcal{S}%
_{q}\rightarrow M_{1}^{+}\left( \Omega ,\Sigma \right) $ which define the
observables of the quantum statistical model. The dual map sends classical
effects to quantum effects; in particular it takes the `crisp' (sharp)
effects represented by characteristic functions to the effects in the range
of the POVM $E^{A}$ associated with $A$:
\begin{equation}
\Sigma \ni X\mapsto A^{\ast }\left( a_{\chi _{X}}\right) =E^{A}\left(
X\right) \,\in \mathcal{E}_{q}\,.  \label{q-dualmap}
\end{equation}
Among the quantum observables $A$ there are injective ones, and the
associated effect valued measures (or POVMs) are called informationally
complete.\footnote{%
The notion of informationally complete observables and first examples are due
to work of the late E. Prugovecki from the 1970s. Examples and a survey of the
early literature on this concept can be found in\cite{OQP}.} If $E^{A}$ is
informationally complete, the dual map $A^{\ast }$ sends the dual space of
$M\left( \Omega ,\Sigma \right) $ onto a dense subspace of the space of bounded
selfadjoint operators (the dual of the space of selfadjoint trace class
operators hosting $\mathcal{S}_{q}$). However, the set of measurable classical
effects which arises as the convex hull of all $a_{\chi
_{X}}$ can never exhaust, or be dense in, the set of quantum effects $%
\mathcal{E}_{q}$: this is due to the fact that for an informationally
complete POVM, the effects $E^{A}\left( X\right) $ cannot be projections
other than $O$ or $I$.\footnote{%
A proof of this fact was given by Busch and Lahti\cite{BuLa89}. Intuitive
demonstrations for the finite dimensional case can be found in\cite
{BuHelStul93}.} This means that an informationally complete observable $A$
does not induce a `good' classical embedding of the full quantum statistical
model $\langle \mathcal{S},\mathcal{E}\rangle $ but only of a reduced
quantum model $\langle \mathcal{S},\mathcal{E}\left( A\right) \rangle $,
with $\mathcal{E}$ replaced by $\mathcal{E}\left( A\right) =E^{A}\left(
\Sigma \right) $, the separating effect algebra consisting of the range (or
the convex hull of the range) of the POVM $E^{A}$.

An example of an informationally complete observable is given by the
coherent state based phase space POVM, which corresponds to the Husimi
distribution functions associated with each quantum state. These are bona
fide phase space probability densities but the price to be paid is that its
marginal position and momentum distributions are convolutions of the
standard quantum mechanical position and momentum distributions with
Gaussian confidence distributions, thus ensuring that appropriate Heisenberg
uncertainty relations for the inaccuracies are satisfied. Wigner's theorem
can thus be interpreted as a demonstration of the incompatibility of the
standard position and momentum observables, while the existence of the
Husimi and other, similar phase space distributions shows that joint
measurements of position and momentum are possible if an allowance is made
for unsharpness in line with the uncertainty relation.\footnote{%
A detailed account of phase space observables, joint measurements of
position and momentum, the role of the Heisenberg uncertainty relation, and
a survey of relevant literature is given in\cite{OQP}.}

Informationally complete observables for $\mathcal{H}=\Bbb{C}^{2}$ are
easily constructed\cite{Bu86}, together with simple, realizable measurement
schemes.\cite{Bu87,OQP} Geometrically, they can be represented as affine
embeddings of the Poincar\'{e} sphere $\widetilde{\mathcal{S}}\left( \Bbb{C}%
^{2}\right) $ into the set $M_{1}^{+}\left( \Omega ,\Sigma \right) $. With
the choice $\Omega _{4}=\left\{ 1,2,3,4\right\} $ one can ensure that the
embedding is not only injective but that its linear extension is surjective.
In this case the sphere $\widetilde{\mathcal{S}}\left( \Bbb{C}^{2}\right) $
is mapped onto an ellipsoid which is embedded into the tetrahedron $\mathcal{%
S}\left( \Omega _{4}\right) $. The dual map sends the hypercube $\mathcal{E}%
\left( \Omega _{4}\right) $ into a `stretched' hypercube inside the
`diamond' $\widetilde{\mathcal{E}}\left( \Bbb{C}^{2}\right) $, in such a way
that the elements $\left( 1,1,1,1\right) $ and $\left( 0,0,0,0\right) $
[which represent the $I$ and $O$ effects] are mapped to $\left( 2,\mathbf{0}%
\right) $ and $\left( 0,\mathbf{0}\right) $, respectively. This description
makes it evident that the extreme points of the set of quantum effects
[other than $I$ and $O$] cannot be represented in terms of classical
effects, except, perhaps, in finitely many cases where some extreme points
of the hypercube touch the extreme boundary of the diamond.

\subsection{Cayley representation of the $\Bbb{C}^{2}$ statistical duality}

The association $\rho \longleftrightarrow \left( 1,\mathbf{r}\right) $, $%
a\longleftrightarrow \left( a_{0},\mathbf{a}\right) $ reviewed above defines
a bijective affine mapping $\Phi :$ $\mathcal{S}\left( \Bbb{C}^{2}\right)
\rightarrow \widetilde{\mathcal{S}}\left( \Bbb{C}^{2}\right) $. Likewise,
the dual map $\Phi ^{\ast }:\widetilde{\mathcal{E}}\left( \Bbb{C}^{2}\right)
\rightarrow \mathcal{E}\left( \Bbb{C}^{2}\right) $ is an affine bijection.
Hence we have both a `good' embedding (via $\Phi $) and a `good' extension
(via $\Phi ^{-1}$) of $\langle \mathcal{S}\left( \Bbb{C}^{2}\right) ,%
\mathcal{E}\left( \Bbb{C}^{2}\right) \rangle $. In fact the linear
extensions of these maps are isometries with respect to the trace norm $%
\left| \left| a\right| \right| _{1}=$\textrm{tr}$\left[ \left| a\right| %
\right] $ and the norm $\left| \left| \left( a_{0},\mathbf{a}\right) \right|
\right| =\mathrm{\max }\left\{ a_{0},\left| \mathbf{a}\right| \right\} $.

\subsection{Gleason's theorem (and a simple variant)}

The essence of \emph{standard} quantum mechanics is captured in the standard
quantum statistical model $\langle \mathcal{S}_{q},\mathcal{E}%
_{q}^{p}\rangle $. Let $v:\mathcal{E}_{q}^{p}\rightarrow \left[ 0,1\right] $
be a generalized probability measure on the lattice of projections $\mathcal{%
E}_{q}^{p}$, that is, a map which satisfies the conditions $v\left( O\right)
=0$, $v\left( I\right) =1$, and $v\left( \sum_{k}P_{k}\right)
=\sum_{k}v\left( P_{k}\right) $ for any finite or countable set of pairwise
orthogonal $P_{k}\in \mathcal{E}_{q}^{p}$. Let $\mathcal{S}_{q}^{p}$ denote
the set of all such generalized probability measures. This is a convex set,
and each projection $P\in \mathcal{E}_{q}^{p}$ defines an effect $a_{P}$ on $%
\mathcal{S}_{q}^{p}$ via $v\mapsto a_{P}\left( v\right) =v\left( P\right) $.
These effects separate $\mathcal{S}_{q}^{p}$ since $v\left( P\right)
=v^{\prime }\left( P\right) $ for all $P$ implies $v=v^{\prime }$. Hence $%
\langle \mathcal{S}_{q}^{p},\mathcal{E}_{q}^{p}\rangle $ is a statistical
model. Gleason's theorem asserts that if the dimension of the underlying
Hilbert space is greater than 2, then for every $v\in \mathcal{S}_{q}^{p}$,
there is a unique $\rho \in \mathcal{S}_{q}$ such that $v\left( P\right) =%
\mathrm{tr}\left[ \rho \cdot P\right] $ for all $P\in \mathcal{E}_{q}^{p}$.
This association $R_{\mathrm{G}}:v\mapsto \rho =\rho _{v}$ is bijective and
thus affine. Hence it induces a `good' embedding \emph{and} extension of $%
\langle \mathcal{S}_{q},\mathcal{E}_{q}^{p}\rangle $ in terms of $\langle
\mathcal{S}_{q}^{p},\mathcal{E}_{q}^{p}\rangle $.

A similar but much simpler result arises for the statistical model $\langle
\mathcal{S}_{q},\mathcal{E}_{q}\rangle $ of $\emph{operational}$ quantum
mechanics if we define $\mathcal{S}_{q}^{e}$ to be the set of all
generalized probability measures on the full set of quantum effects $%
\mathcal{E}_{q}$. The set $\mathcal{E}_{q}$ comprises enough elements so as
to include basis systems of effects $a_{k}$ such that $\sum_{k}a_{k}$ is an
effect. This readily entails that every generalized probability measure $v$
on $\mathcal{E}_{q}$ extends to a bounded linear functional on the space of
bounded selfadjoint operators. The $\sigma $-additivity of $v$ implies that
this functional arises from a density operator via the trace formula, $%
v\left( a\right) =\mathrm{tr}\left[ \rho \cdot a\right] $. Hence we have a
`good' embedding and extension of $\langle \mathcal{S}_{q},\mathcal{E}%
_{q}\rangle $ to a statistical model $\langle \mathcal{S}_{q}^{e},\mathcal{E}%
_{q}\rangle $.\footnote{%
This fact is proved and discussed for quantum mechanics in \cite{Bu03}. It
has been proved in a much more abstract context by Beltrametti and Bugajski
\cite{Bug2}.} This time there is no restriction on the dimension of the
Hilbert space. The generalized probability measures on the set of effects
restrict to generalized probability measures on the projections, which
entails that $\mathcal{S}_{q}^{e}$ is a proper subset of $\mathcal{S}_{q}^{p}
$ in the case $\mathcal{H}=\Bbb{C}^{2}$ while $\mathcal{S}_{q}^{e}=\mathcal{S%
}_{q}^{p}$ in the case $\dim \mathcal{H}>2$.

\subsection{Compound systems extension}

Let $\widetilde{\mathcal{H}}=\mathcal{H}\otimes \mathcal{H}^{\prime }$ be
the tensor product of the Hilbert space $\mathcal{H}$ of the system of
interest with an auxiliary Hilbert space $\mathcal{H}^{\prime }$. Denote by $%
R$ the partial trace map from the trace class of $\widetilde{\mathcal{H}}$ onto
the trace class of $\mathcal{H}$. This is a linear map, and its restriction to
$\mathcal{S}( \widetilde{\mathcal{H}}) $ has as
its range all of $\mathcal{S}\left( \mathcal{H}\right) $. The dual map $%
R^{\ast }:a\mapsto a\otimes I$ is a linear injection or all bounded
operators of $\mathcal{H}$ into the space of bounded linear selfadjoint
operators of $\widetilde{\mathcal{H}}$, with the property $O\leq a\leq I$ $%
\Rightarrow O\leq R^{\ast }\left( a\right) \leq R^{\ast }\left( I\right)
=I\otimes I$. In this way, the statistical model $\langle \mathcal{S}(
\widetilde{\mathcal{H}}) ,\mathcal{E}( \widetilde{\mathcal{H}}%
) \rangle $ arises as a `good' extension of the statistical model $%
\langle \mathcal{S}\left( \mathcal{H}\right) ,\mathcal{E}\left( \mathcal{H}%
\right) \rangle $.

\subsection{Canonical classical extension of quantum mechanics}

We now assume that we are given a `good' classical extension induced by a
surjective affine map $R:M_{1}^{+}\left( \Omega ,\Sigma \right) \rightarrow
\mathcal{S}\left( \mathcal{H}\right) $. We will make a few assumptions on $%
\left( \Omega ,\Sigma \right) $ and take a few steps to eliminate
`redundancies'. This will lead us in a fairly natural way to exhibit the
canonical classical extension introduced by Misra and recognized by Bugajski
as a `good' extension of the maximal quantum statistical model into a
distinguished classical statistical model. The assumptions on $\left( \Omega
,\Sigma \right) $ are:\newline
(a) $\Sigma $ contains all singletons $\left\{ \omega \right\} $, $\omega
\in \Omega $;\newline
(b) the extreme elements of $M_{1}^{+}\left( \Omega ,\Sigma \right) $ are
exactly the Dirac measures $\delta _{\omega }$, so that each $\mu \in
M_{1}^{+}\left( \Omega ,\Sigma \right) $ can be uniquely written as
\begin{equation}
\mu =\int_{\Omega }\delta _{\omega }\,d\mu \left( \omega \right) .
\label{mu-convdec}
\end{equation}

Consider any density operator $\rho $ corresponding to a pure state, $\rho
=P_{\varphi }\equiv |\varphi \rangle \langle \varphi |$. Let $\mu \in
M_{1}^{+}\left( \Omega ,\Sigma \right) $ be any measure for which $R\mu
=\rho $. It follows that $R\delta _{\omega }=\rho $ for all $\delta _{\omega
}$ which occur in the convex decomposition (\ref{mu-convdec}) of $\mu $, and
every pure state will then be an image under $R$ of some $\delta _{\omega }$%
. We will consider $\Omega $ as restricted to those $\omega $ for which $%
R\delta _{\omega }$ is pure. [Note that this presupposes that $\left\{
\omega \in \Omega :R\delta _{\omega }\;\mathrm{is\;pure}\right\} \in \Sigma $%
. We will see presently that this can be trivially guaranteed.] Further we
can identify (as equivalent) all $\omega ,\omega ^{\prime }\in \Omega $ for
which $R\delta _{\omega }=R\delta _{\omega ^{\prime }}$, and replace $\Omega
$ with the set (also denoted $\Omega $) of all equivalence classes $\left[
\omega \right] $, $\Sigma $ with the $\sigma $-algebra (again denoted $%
\Sigma $) of sets $\left[ X\right] :=\left\{ \left[ \omega \right] :\omega
\in X\right\} $ and $R$ with the induced map, also called $R$, that sends $%
\delta _{\left[ \omega \right] }$ to $R\delta _{\omega }$. In this way we
identify $\Omega $ with \textrm{Ex}$\left[ \mathcal{S}\left( \mathcal{H}%
\right) \right] $, the set of pure density operators.

Now we let $\Omega =\mathrm{Ex}\left[ \mathcal{S}\left( \mathcal{H}\right) %
\right] $ and $\omega $ denote any pure state in $\mathcal{S}\left( \mathcal{%
H}\right) $. The affine reduction map $R$ must then satisfy
\begin{equation}
R:\mu =\int_{\Omega }\delta _{\omega }\,d\mu \left( \omega \right) \mapsto
\int_{\Omega }R\delta _{\omega }\,d\mu \left( \omega \right) =\int_{\Omega
}\omega \,d\mu \left( \omega \right) =\rho _{\mu }\,.  \label{Misramap1}
\end{equation}
The nondegenerate bilinear forms induced by the classical and quantum
statistical models involved is given by
\begin{equation}
\langle \rho _{\mu },a\rangle =\mathrm{tr}\left[ \rho _{\mu }\cdot a\right]
=\int_{\Omega }\mathrm{tr}\left[ \omega \cdot a\right] \,d\mu \left( \omega
\right) =\langle \mu ,f_{a}\rangle \,.  \label{canon-bilinform}
\end{equation}
This determines the dual map
\begin{equation}
R^{\ast }:a\mapsto f_{a}\,,\;\;\;f_{a}\left( \omega \right) =\mathrm{tr}%
\left[ \omega \cdot a\right] \,.  \label{Misramap2}
\end{equation}
For this formula to make precise sense, it is necessary that $\Sigma $ is
fixed in such a way that the functions $f_{a}$ are measurable. Misra\cite
{Misra} has shown how to achieve this. Thus every quantum effect $a\in
\mathcal{E}\left( \mathcal{H}\right) $ is represented as a classical
measurable effect $f_{a}$. Hence $R$ induces a `good' classical extension of
the quantum statistical duality.\footnote{%
The Misra-Bugajski map is an instance of a much more general mathematical
result of Choquet on boundary measures of compact convex sets in the context
of locally convex vector spaces. See Theorem I..4.8 of Alfsen\cite{Alf}.}

We can see from Eq. (\ref{Misramap2}) that all quantum effects, including
the `sharp' or `crisp' ones given by projections, are represented by
functions whose values vary continuously between their maximum and minimum
values. For any projection $a$ not equal to $I$ or $O$, $f_{a}\left( \omega
\right) $ assumes all values in $\left[ 0,1\right] $. Hence the projections
are represented as classical fuzzy sets. All quantum observable, represented
by POVMs $E$, including the PVMs, are described under the map $R^{\ast }$ as
classical fuzzy random variables with associated Markov kernel $K^{E}\left(
\omega ,X\right) =\mathrm{tr}\left[ \omega \cdot E\left( X\right) \right] \,$%
.

The many-to-one relationship between the probability measures $\mu $ and the
quantum states $\rho _{\mu }=R\mu $ reflects exactly the infinitely many
ways in which every density operator which is not a pure state can be
written as a convex combination of pure states. The set of all classical
effect valued measures $R^{\ast }\circ E$ represents all quantum mechanical
measurements as classical fuzzy measurements which even taken together are
too imprecise to separate the various different probability measures $\mu
,\mu ^{\prime },\dots $ which lead to the same density operator $\rho =R\mu
=R\mu ^{\prime }=\dots $.

\section{Concluding remarks}

In this contribution I have reviewed the concept of statistical model and the
various possible relations of embeddings and extensions between statistical
models. The possible ways of constructing `good' classical representations of
either the full or some restricted quantum statistical models have been
exhibited. The only `good' \emph{classical extension} of the full quantum
statistical model is uniquely given (modulo redundancies) by the Misra-Bugajski
map. This canonical classical extension of quantum mechanics does not
constitute a hidden-variable completion of quantum mechanics because it does
\emph{not} render the sharp quantum properties (projections) as dispersion-free
in the extremal classical states. On the contrary, under this classical
extension \emph{all} quantum effects -- whether unsharp or sharp -- are
represented as classical fuzzy sets.

While the classical embeddings of quantum mechanical statistical models via
effect valued measures (which may or may not be informationally complete)
have been extensively studied and led to a plethora of applications\cite{OQP}%
, much remains to be done to explore and exhaust the full potential of the
canonical classical embeddings. In their very last joint work, Beltrametti and
Bugajski made some tentative steps to characterize quantum correlations as
opposed to classical correlations in this framework. The `Bell phenomenon in
classical framework', that is, the violation of Bell's inequalities for fuzzy
random variables corresponding to EPR-Bell observables, calls for an
investigation of the notion of coexistence (joint measurability) of fuzzy
random variables. More generally, it would be desirable to cast all the
distinctive quantum structures, such as non-commutativity, complementarity,
uncertainty relations, entanglement, in the language of the classical canonical
extension. This should enable us to relate these quantum features to concepts
closer to our `classical' experience and intuition -- in other words: it should
contribute to a better understanding of quantum mechanics.

\end{document}